\documentclass{article}
\usepackage{graphicx} 
\usepackage{tikz} 
\usetikzlibrary{arrows,automata, positioning} 
\usepackage{amsthm} 
\usepackage{amsfonts} 
\usepackage{amsmath} 
\usepackage[linesnumbered,ruled,vlined]{algorithm2e} 
\usepackage{float}
\usepackage{hyperref}
\usepackage{pdfpages}
\newtheorem{theorem}{Theorem}[section]
\newtheorem{corollary}{Corollary}[theorem]
\newtheorem{lemma}[theorem]{Lemma}
\theoremstyle{definition}
\newtheorem{definition}{Definition}[section]
\newcommand{\diam}{\mathrm{diam}}

\usepackage{pgfplots}

\title{Monitoring graph edges via shortest paths: computational complexity and approximation algorithms}

\date{December 2023}

\begin{document}

    \begin{figure}[H]
    \centerline{\includegraphics[scale=0.65,keepaspectratio]{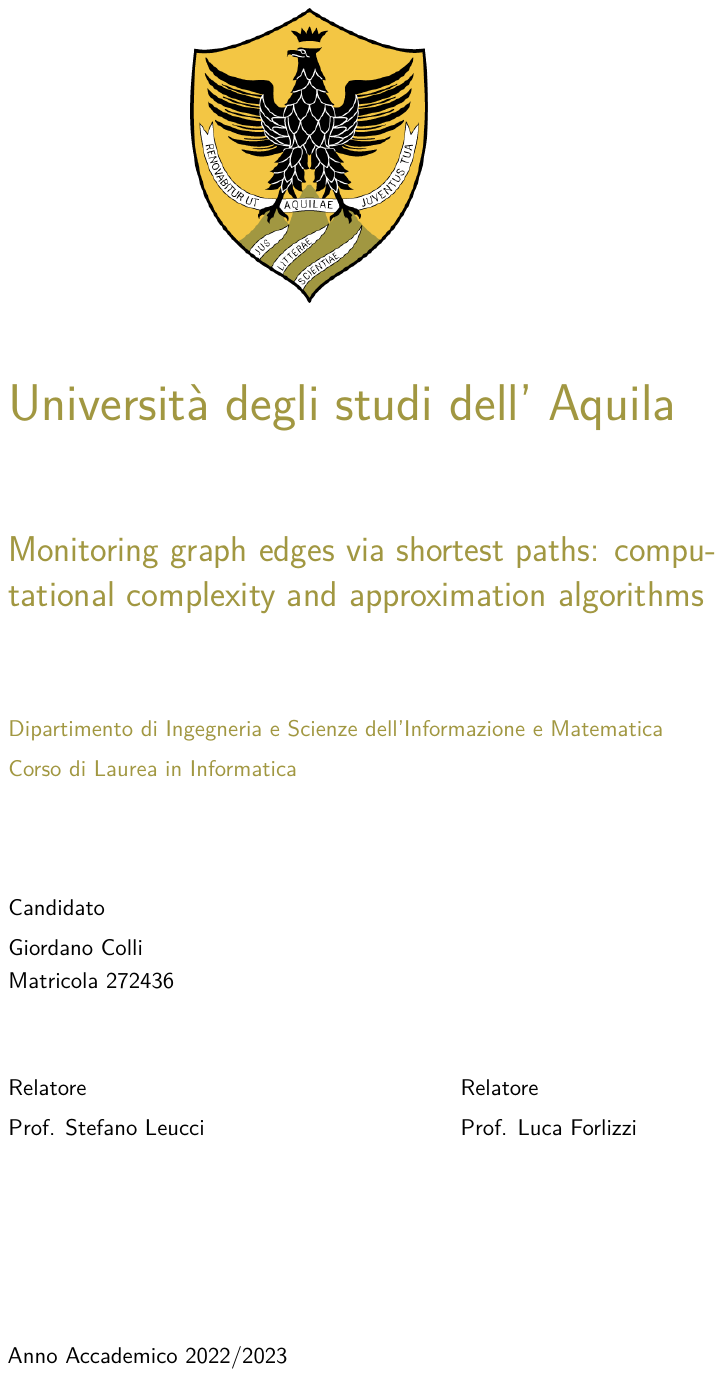}}
    \end{figure}

\maketitle


\begin{center}
    \textbf{Abstract}
\end{center}
Edge-Geodetic Sets \cite{HASLEGRAVE202379} play a crucial role in network monitoring and optimization, wherein the goal is to strategically place monitoring stations on vertices of a network, represented as a graph, to ensure complete coverage of edges and mitigate faults by monitoring lines of communication. This paper illustrates and explores the Monitoring Edge-Geodetic Set (\textsc{MEG-set}) problem, which involves determining the minimum set of vertices that need to be monitored to achieve geodetic coverage for a given network. The significance of this problem lies in its potential to facilitate efficient network monitoring, enhancing the overall reliability and performance of various applications.
In this work, we prove the $\mathcal{NP}$-completeness of the \textsc{MEG-set} optimization problem by showing a reduction from the well-known \textsc{Vertex Cover} problem. Furthermore, we present inapproximability results, proving that the \textsc{MEG-set} optimization problem is $\mathcal{APX}$-Hard and that, if the unique games conjecture holds, the problem is not approximable within a factor of $2-\epsilon$ for any constant $\epsilon > 0$. Despite its $\mathcal{NP}$-hardness, we propose an efficient approximation algorithm achieving an approximation ratio of $O(\sqrt{|V(G)| \cdot \ln{|V(G)|})}$ for the \textsc{MEG-set} optimization problem, based on the well-known \textsc{Set Cover} approximation algorithm,
where $|V(G)|$ is the number of nodes of the \textsc{MEG-set} instance. These results shed light on the complexity of the \textsc{MEG-set} optimization problem and provide valuable insights into the challenges of achieving near-optimal solutions for practical network monitoring scenarios.

\newpage
\section{Introduction}
We study the \emph{Monitoring Edge-Geodesic Set} (\textsc{MEG-set}) problem, which arises from the context of network monitoring challenges, specifically in identifying and addressing faults within networks. Our networks are modeled as finite, undirected, simple and connected graphs, with vertices representing hosts and edges denoting connections. The goal is to promptly identify network \textbf{edge failures} through changes in distance between probe pairs. In the realm of network monitoring, the objective is to detect faults using distance probes. To achieve this, we strategically select a subset of vertices so that the shortest paths between pairs of vertices cover all network connections.

\noindent The \textsc{MEG-set} problem was introduced in \cite{10.1007/978-3-031-25211-2_19}. In their paper, the authors have extensively examined the minimum size of \textsc{MEG-set}s across various fundamental graph classes. Specifically, these investigations encompass graph types such as trees, cycles, unicyclic graphs, complete graphs, grids, hypercubes, and corona products.
\noindent The \textsc{MEG-set} \textit{Decision Problem} was also studied in \cite{HASLEGRAVE202379}, where the author shows that the problem is $\mathcal{NP}$-Complete and also gives the best possible upper and lower bounds for the Cartesian and strong products of two graphs. These bounds establish the exact value in many cases, including many examples of graphs whose only Monitoring Edge-Geodetic Set is the whole vertex set.

\noindent Furthermore, in \cite{dev2023monitoring} it was proved that computing the smallest \textsc{MEG-set} size for a graph is $\mathcal{NP}$-Hard even for graphs with a maximum degree of at most $9$.

\noindent In this thesis we have established a compelling relationship between the \textsc{Vertex Cover} \textit{Optimization Problem} and the Monitoring Edge-Geodetic Set (\textsc{MEG-set}) \textit{Optimization Problem} through a carefully devised graph construction. In this construction, we demonstrate how instances of the \textsc{Vertex Cover} \textit{Optimization Problem} can be effectively translated into instances of the \textsc{MEG-set} \textit{Optimization Problem}.

\noindent Specifically, the graph construction highlights that finding a solution to the \textsc{MEG-set} \textit{Optimization Problem} enables us to derive a solution for the \textsc{Vertex Cover} \textit{Optimization Problem}. This connection underlines the inherent computational intricacies shared between the two problems. We employ this connection to prove that the \textsc{MEG-set} \textit{Optimization Problem} is $\mathcal{APX}$-Hard. Moreover, assuming the validity of the unique games conjecture (\cite{10.1145/509907.510017}), we utilize this relationship to establish that $(2-\epsilon)$-approximation algorithm for the \textsc{MEG-set} \textit{Optimization Problem}, where $\epsilon$ is a positive constant, does not exist.

\noindent Our investigations contributes not only to the understanding of \textsc{MEG-set} but also to the broader exploration of connections between different graph-theoretic problems. This newfound insight provides a foundation for further investigations into the computational complexity and interdependencies of problems in the realm of network monitoring and graph theory.

\noindent Furthermore, despite the inapproximability results highlighted by the reduction from the \textsc{Vertex Cover} \textit{Optimization Problem}, we contribute to the practical applicability of Monitoring Edge-Geodetic Sets by presenting an effective approximation algorithm. Our algorithm achieves an approximation factor of $\sqrt{|V(G)| \cdot \ln{|V(G)|}}$, where $|V(G)|$ represents the size of the set of vertices in the input graph $G$.

\noindent This approximation algorithm, which relies on the well-known \textsc{Set Cover} greedy approximation algorithm, provides a valuable balance between theoretical complexity and practical utility, allowing for the selection of vertices in network monitoring scenarios. By introducing this algorithm, we offer a constructive approach to address the challenges associated with computing Monitoring Edge-Geodetic Sets.
\newpage
\section{Preliminaries}
    In this section, there are some preliminary definitions used throughout the following chapters.
    \begin{itemize}
        \item When referring to a ``pair'', the term denotes an unordered pair. 
        \item By $[A]^k$ we denote the set of all $k$-element subsets of $A$, where $A$ is a finite set.
        \item $G=(V, E)$ denotes a simple, undirected, and connected graph.
        \item $V(G)$ is a function that takes in input a graph $G$ and returns the set containing all the vertices of $G$.
        \item $E(G)$ is a function that takes in input a graph $G$ and returns the set containing all the edges of $G$.
        \item The degree of a vertex, say $\deg(v)$, is the number of edges incident to $v$. More formally $\deg(v) = |\{\{u,v\} : \{u,v\} \in E(G)\}|$.
        \item The diameter of a graph $G$, say $\diam(G)$, is the length of the longest shortest path between any two vertices.
        \item A path in $G$ is a tuple $P = \langle e_1,e_2, ..., e_n \rangle$ such that $e_i \in E(G), \forall i \in \mathbb{N}: 1 \le i \le n$, and consecutive edges in the tuple share a common vertex, i.e., $e_i = \{v_{i-1},v_{i}\} \quad \forall i \in \mathbb{N}: 2 \le i \le n$. 
        \item The length of a path $P$ is the size of $P$, i.e., the number of edges contained in $P$.
        \item Distance between $2$ vertices $x,y \in V(G)$ in a graph $G$, namely $d_{G}(x,y)$, is the length of a shortest path between $x$ and $y$ in $G$. 
        \item A problem $\pi \subseteq I_{\pi} \times S_{\pi}$ is a relation, where $I_{\pi}$ is the set of input instances and $S_{\pi}$ is the set of the solutions of the problem.
        \item A Decision problem is that of checking if a given property holds for any input $x \in I_{\pi}$.
                \begin{itemize}
                    \item $S_{\pi}$ = $\{$true,false$\}$.
                    \item $\pi \subseteq I_{\pi} \times S_{\pi}$ corresponds to a function $f:I_{\pi} \to \{$true,false$\}$.
                \end{itemize}
        \newpage
        \item An Optimization problem $\pi$ is a quadruple $\langle I_{\pi}, S_{\pi}, m_{\pi}, goal_{\pi}\rangle$
        \begin{itemize}
            \item $I_{\pi}$ is the set of input instances.
            \item $S_{\pi}$ is the set of feasible solutions for $x \in I_{\pi}$.
            \item $m_{\pi}(x,y)$ is the objective function, i.e., the value of the feasible solution $y \in S_{\pi}$ for $x \in I_{\pi}$.
            \item $goal_{\pi} \in \{min, max\}$ specifies if the objective function must be minimized or maximized.
        \end{itemize}
        \item TIME$(f(n))$ is the class of decision problems \textbf{deterministically} solvable in a time complexity bounded by $O(f(n))$.
        \item NTIME$(f(n))$ is the class of decision problems  \textbf{non-deterministically} solvable in a time complexity bounded by $O(f(n))$ .
        \item $P$ is the class of decision problems \textbf{deterministically} solvable in \textbf{polynomial time}
            \[P = \bigcup_{i=0}^{\infty} \text{TIME}(n^i)\]
        \item $\mathcal{NP}$ is the class of decision problems \textbf{non-deterministically} solvable in \textbf{polynomial time}
            \[NP = \bigcup_{i=0}^{\infty} \text{NTIME}(n^i)\]
        \item $\mathcal{NP}$-Hard is a class of decision problems. A decision problem, say $H$, is $\mathcal{NP}$-Hard if there is a polynomial-time reduction from each problem in $\mathcal{NP}$ to $H$.
        \item $\mathcal{NP}$-Complete is the class of decision problems that are at the same time in $\mathcal{NP}$ and in $\mathcal{NP}$-Hard.
        \item An $r$-approximation algorithm for a minimization problem is an algorithm that, in polynomial time, finds a solution that is within a factor $r$ of the optimal solution. More formally let $O_{opt}$ be the optimal solution to a minimization problem, and let $O_{alg}$ be the solution found by the algorithm. The algorithm is called an $r$-approximation algorithm if, for any instance of the problem: $O_{alg} \le r \cdot O_{opt}$.
        \item An optimization problem, say $p$, is in $\mathcal{APX}$-Hard if there exists a positive constant $\delta > 0$ such that for any $r$-approximation algorithm that solves $p$, $r \ge (1 + \delta)$.
        
    \end{itemize}

\section{Problem Definition}
In this chapter, we provide a formal definition of the problem studied in this thesis. The problem was originally introduced in \cite{dev2023monitoring}.
The Monitoring Edge-Geodetic Set Optimization problem in a graph can be thought of as finding the most strategic and efficient locations (vertices) in a network so that every edge is monitored by at least a pair of selected vertices. A solution consists of the fewest critical locations in a graph that, when monitored, provide complete coverage of the entire network.\\

\begin{definition} 
    An edge $\{u,v\} \in E(G)$ is \textit{monitored} by a pair of vertices $\{x,y\}$ with $x,y \in V(G)$ if $\{u,v\}$ belongs to all shortest paths from $x$ to $y$ in $G$.
\end{definition}

\begin{definition}[\textsc{MEG-set}]
    A Monitoring Edge-Geodetic Set of $G$ is a subset $M \subseteq V(G)$ of vertices such that $\forall \{u,v\} \in E(G) \quad \exists x,y \in M $ such that the pair $\{x,y\}$ monitors the edge $\{u,v\}$.
\end{definition}

\begin{definition}[\textsc{MEG-set} Decision Problem]
Given a graph $G$ and $k \in \mathbb{N}$, 
the Monitoring Edge-Geodetic decision problem is that of deciding whether there exists a \textsc{MEG-set} of $G$ containing at most $k$ vertices.
\end{definition}
\begin{definition}[\textsc{MEG-set} Optimization Problem]
Given a graph $G$,\\ the Monitoring Edge-Geodetic optimization problem is that of finding a \textsc{MEG-set} of $G$ with minimum size. 
\end{definition}

\noindent In the following we will prove that \textsc{MEG-set} \textit{Decision} Problem is $\mathcal{NP}$-Complete, thus implying that the \textit{Optimization} version is $\mathcal{NP}$-Hard. \\
Notice that the $\mathcal{NP}$-hardness of the problem was already proved in the paper \cite{HASLEGRAVE202379}, but our reduction will also be instrumental in showing that the optimization version is $\mathcal{APX}$-hard.
\newpage

\section{$\mathcal{NP}$-Completeness} 
To prove the $\mathcal{NP}$-Hardness of this problem, we devise a polynomial-time reduction from the well-known \textsc{Vertex Cover} problem. This reduction constructs an instance of the \textsc{MEG-set} problem from an instance of \textsc{Vertex Cover}. By leveraging this reduction, we establish a relation between the sizes of the two optimum solutions. We now give some useful lemmas about the basic properties of a \textsc{MEG-set}.
\begin{lemma}[Leaf Lemma]
    \label{leafLemma}
    Let $v \in V(G)$ be a leaf \footnote{A leaf vertex of a graph $G$ is a vertex $u \in V(G)$ such that $ \deg(u) = 1$.}. $v $ belongs to all \textsc{MEG-set}s of $G$.
    \begin{proof}
        Let $u$ be the sole neighbor of $v$ (\autoref{leafLemmaPicture}).
        Suppose towards a contradiction that $v \not \in M$. Since $\{u,v\}$ is monitored by a pair of vertices $\{x,y\}$ with $x,y \in M$, $v$ must be an internal vertex of all shortest paths from $x$ to $y$.  
        This implies that $v$ has at least two incident edges, which contradicts the fact that $v$ is a leaf.

    \end{proof}
    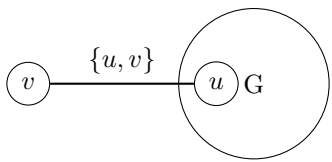
\begin{figure}[H]
        \centering
        
        \begin{tikzpicture}
                \node[shape=circle,draw=black] (v) at (0,0) {$v$};
                \node[shape=circle,draw=black,minimum size=2cm] (G) at (3,0) {G};
                \node[shape=circle,draw=black] (u) at (2.5,0) {$u$};
                \path [-,thick] (v) edge node[above] {$\{u,v\}$} (u);
        \end{tikzpicture}
        \caption{$u$ is the sole neighbor of $v$.}
        \label{leafLemmaPicture}
    \end{figure}
        
\end{lemma}

\begin{lemma}
    \label{superiorLeafLemma}
    Let $\{u,v\} \in E(G)$ be an edge such that $v$ is a leaf and $u$ is not a leaf, and let $M$ be a \textsc{MEG-set} of $G$. $M \setminus \{u\}$ is a \textsc{MEG-set} of $G$.
    \begin{proof}
        We prove that $M \setminus \{u\}$ is a \textsc{MEG-set} by showing that each edge $\{a,b\} \in E(G) $ is monitored by some pair $\{x,y\}$ with $ x,y \in M \setminus \{u\}$. \\
        Note that $v \in M$ due to Lemma \ref{leafLemma}.
        We split the proof into 2 cases: 
        \begin{itemize}
            \item Case 1: $\{a,b\} \neq \{u,v\}$. If the edge $\{a,b\}$ is monitored by a pair $\{u,y\}$ then $y \neq v$ (since the unique shortest path from $u$ to $v$ contains only the edge $\{u,v\}$) and $\{a,b\}$ is also monitored
by the pair $\{v,y\}$, where $ v,y \in M \setminus \{u\}$. 
            \item Case 2: $\{a,b\} = \{u,v\}$.  A vertex $w: w \not\in \{u,v\}$ must belong to the \textsc{MEG-set} $M$ because $u$ has a degree greater than or equal to two. That implies $\{a,b\}$ is monitored by $\{v,w\}$.
        \end{itemize}
\end{proof}
\end{lemma}

\newpage
\noindent We show a reduction from \textsc{Vertex Cover} to \textsc{MEG-set}. We focus on instances of \textsc{Vertex Cover} on graphs with diameter smaller than or equal to two. The \textsc{Vertex Cover} \textit{Decision} Problem remains $\mathcal{NP}$-Hard also in this case. \footnote{\cite{10.1145/800119.803884} considers the \textsc{Vertex Cover} problem with no restriction on the diameter of the input graph $G$. However the hardness extends easily to graphs of diameter at most $2$ by noticing that, whenever $G$ is not a clique, the size of a minimum vertex cover of the graph obtained by adding a universal vertex to $G$ is exactly that of a minimum vertex cover of $G$ plus $1$.} This reduction will be used later in this paper to prove the $\mathcal{APX}-$Hardness of the \textsc{MEG-set} \textit{Optimization} Problem.
\subsection{Reduction from \textsc{Vertex Cover}}
    \label{construction}
    Let $H = (V_H,E_H)$ be an input graph with $|V(H)| \ge 2$ and $\diam(H) \le 2$. \\ 
    We construct a new graph $G$ by starting from $H$ and augmenting it as follows: for each vertex $v \in V(H)$ we add 2 vertices $v'$ and $v''$ to $H$, then connect $v$ to $v'$ and $v'$ to $v''$. We denote the set containing all the vertices $v'$ as $L'$ and the set containing all the vertices $v''$ as $L''$. Notice that: \begin{enumerate}
        \item[i)] Each $v'' \in L''$ belongs to all \textsc{MEG-set}s of $G$ due to Lemma \ref{leafLemma}
        \item[ii)] Each $v' \in L'$ does not belong to any optimal \textsc{MEG-set} of $G$ due to Lemma \ref{superiorLeafLemma}
    \end{enumerate}
    The basic idea is to monitor each edge that is incident to a vertex $v \in V(H)$ that belongs to a selected \textsc{MEG-set}. We denote the set of edges $\{v,v'\}$ as $E_{L'}$ and the set of edges $\{v',v'' \}$ as $E_{L''}$.  Note that at this point of the construction, all edges are monitored by at least one pair of leaves in $L''$. To avoid this fact we shortcut all the shortest paths from one leaf to another: we insert two vertices $v_{*}$ and $v_{*}'$ and connect them to each other, then we connect $v_{*}$ to all vertices $v' \in L'$. We denote the set containing all the edges $\{v_*,v'\}$, $v' \in L'$ as $E_*$ . The vertex $v_{*}$ is used to create shortcuts between all shortest paths from a vertex in $L' \cup L''$ to a vertex in $L' \cup L''$. Notice that since $v_{*}'$ is a leaf it belongs to all \textsc{MEG-set}s of $G$ due to the Lemma \ref{leafLemma}. The vertex $v_{*}'$ is used to monitor all the edges in $E_* \cup \{v_{*},v_{*}'\}$.  \\~\\
    \noindent Figure \ref{constructionPicture} provides an example of construction of the graph $G$.

    \noindent More formally let $G = (V(H) \cup L' \cup L'' \cup \{v_*,v_{*}'\},E(H) \cup E_{L'} \cup E_{L''} \cup E_* \cup \{ \;  \{v_*,v_{*}'\} \; \}) $, where 
    \begin{itemize}
        \item $L' = \{v': v \in V(H)\}$
        \item $L'' = \{v'': v \in V(H)\}$
        \item $E_{L'} = \{\{v,v'\}: v \in V(H)\}$
        \item $E_{L''} = \{\{v'',v'\}:  v \in V(H)\}$
        \item $E_{*} = \{\{v_*,v'\}:  v \in V(H)\}$
    \end{itemize}
    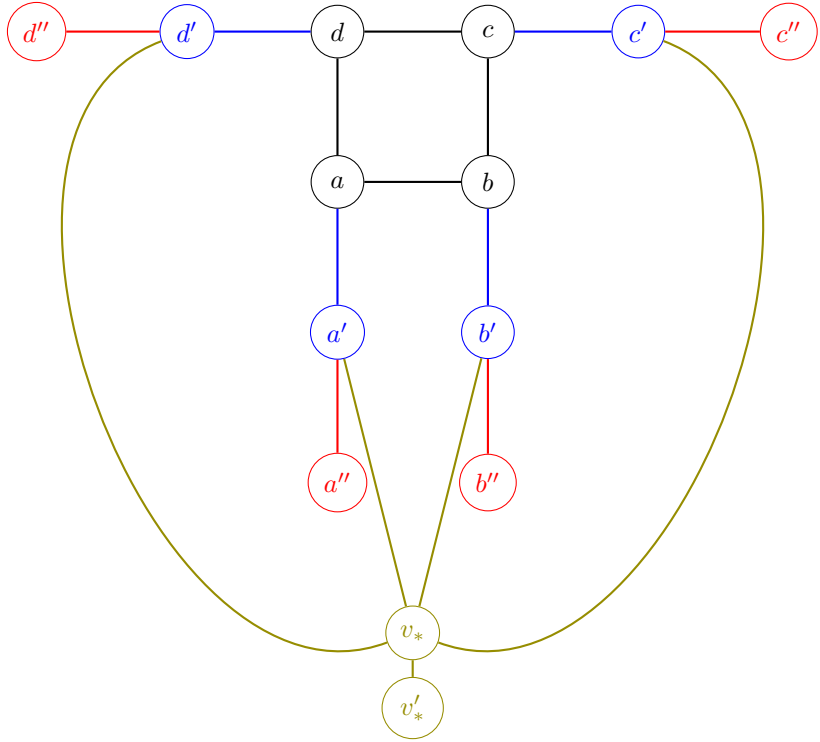
\begin{figure}[H]
        \centering
        
        \begin{tikzpicture}            
                \node[shape=circle,draw=black,minimum size = 0.7cm] (a) at (0,0) {$a$};
                \node[shape=circle,draw=black,minimum size = 0.7cm] (b) at (2,0) {$b$};
                \node[shape=circle,draw=black,minimum size = 0.7cm] (c) at (2,2) {$c$};
                \node[shape=circle,draw=black,minimum size = 0.7cm] (d) at (0,2) {$d$};
                \node[shape=circle,draw=black,minimum size = 0.7cm,blue] (a') at (0,-2) {$a'$};
                \node[shape=circle,draw=black,minimum size = 0.7cm,blue] (b') at (2,-2) {$b'$};
                \node[shape=circle,draw=black,minimum size = 0.7cm,blue] (c') at (4,2) {$c'$};
                \node[shape=circle,draw=black,minimum size = 0.7cm,blue] (d') at (-2,2) {$d'$};
                \node[shape=circle,draw=black,minimum size = 0.7cm,red] (a'') at (0,-4) {$a''$};
                \node[shape=circle,draw=black,minimum size = 0.7cm,red] (b'') at (2,-4) {$b''$};
                \node[shape=circle,draw=black,minimum size = 0.7cm,red] (c'') at (6,2) {$c''$};
                \node[shape=circle,draw=black,minimum size = 0.7cm,red] (d'') at (-4,2) {$d''$};
                \node[shape=circle,draw=black,minimum size = 0.7cm,olive] (*) at (1,-6) {$v_*$};
                \node[shape=circle,draw=black,minimum size = 0.7cm,olive] (*') at (1,-7) {$v_*'$};
                \path [-,thick] (a) edge (b);
                \path [-,thick] (b) edge (c);
                \path [-,thick] (c) edge (d);
                \path [-,thick] (d) edge (a);
                \path [-,thick,blue] (a) edge (a');
                \path [-,thick,blue] (b) edge (b');
                \path [-,thick,blue] (c) edge (c');
                \path [-,thick,blue] (d) edge (d');
                \path [-,thick,red] (a') edge (a'');
                \path [-,thick,red] (b') edge (b'');
                \path [-,thick,red] (c') edge (c'');
                \path [-,thick,red] (d') edge (d'');
                \path [-,thick,olive,>=stealth, bend left=90] (c') edge (*);
                \path [-,thick,olive,>=stealth, bend right=90] (d') edge (*);
                \path [-,thick,olive] (a') edge (*);
                \path [-,thick,olive] (b') edge (*);
                \path [-,thick,olive] (*') edge (*);
        \end{tikzpicture}
        \caption{An example of construction of the graph $G$ from the input graph $H = (\{a,b,c,d\},\{ \; \{a,b\}, \{b,c\}, \{c,d\}, \{d,a\} \; \})$. Vertices and edges in the original graph $H$ are colored in black, Vertices in $L'$ and edges in $E_{L'}$ are colored in \textcolor{blue}{blue}, Vertices in $L''$ and edges in $E_{L''}$ are colored in \textcolor{red}{red}, Vertices in $\{v_{*}',v_{*}\}$ and edges in $E_{*}$ are colored in \textcolor{olive}{olive}.}
        \label{constructionPicture}
    \end{figure}

\newpage
\noindent For the rest of this section, $G$ represents the graph obtained from $H$ according to the construction above.\\
In order to prove the $\mathcal{NP}$-Completeness of the problem, we give two useful preliminary lemmas. The first one shows that if a \textsc{MEG-set} of the graph $G$ having size $k+|V(H)|+1$ is given, then it is possible to obtain a \textsc{Vertex Cover} of $H$ having size $k$. Vice versa, it is possible to construct a \textsc{MEG-set} of $G$ starting from a \textsc{Vertex Cover} of $H$.
\begin{lemma}
    \label{vertexCoverToMeg}
    Let $VC$ be a \textsc{Vertex Cover} of $H$. $M = VC \cup L'' \cup \{v_*'\} $  is a \textsc{MEG-set} of G. Furthermore, let $VC$ be a \textsc{Vertex Cover} of H with size $k$. $M = VC \cup L'' \cup \{v_{*}'\}$ is a \textsc{MEG-set} of $G$ with size $k + |V(H)| + 1$.

    \begin{proof}
        We prove that each edge $e \in E(G)$ is monitored. We distinguish four cases:
        \begin{itemize}
            \item Case 1: $ e \in E(H)$. Let $e = \{u,v\}$. Since $VC$ is a \textsc{Vertex Cover}, at least one of $u$ and $v$ must belong to $VC$, therefore we assume w.l.o.g.\ that $u \in VC$. Each path from $v''$ to  $u$ must either pass through the vertex $v$ or pass through the vertex $v_*$. The shortest path including $v$ is $\langle \{v'',v'\}, \{v',v\}, \{v,u\} \rangle$ and has length $3$.  Each path passing through $v_*$ must contain at least four edges. Therefore $\{u,v\}$ is monitored by $\{v'',u\}$, where both $v''$ and $u$ are in $M$ since $v'' \in L''$ and $u \in VC$.

            \item Case 2:  $e \in E_{L''} \cup E_{*}$. In this case, there is \textit{exactly} one endpoint $u'$ of $e$ that belongs to $L'$ and either $e = \{u'', u'\}$ or $e = \{v_*, u'\}$. Notice that there is a unique shortest path from $v_*'$ to $u''$ including the edges $\{v_*',v_*\}\{v_*,u'\},\{u',u''\}$. This implies that the edge $e$ is monitored by $\{v'_*, u''\}$, where both $v'_*$ and $u'' \in L''$ are in $M$.
            \item Case 3: $e \in E_{L'}$. Let $e = \{u,u'\}$ where $u$ is the unique endpoint of $e$ in $V(H)$. If $u \in VC$ then there is a single shortest path from $u''$ to $u$ and it includes the edge $\{u,u'\}$.  \\ 
            If $u \not\in VC$ then there exists a neighbor $v$ of $u$ such that $v \in VC$. Similar arguments to those used in Case 1 show that there is a single shortest path from $u'' \in L''$ to $v$ containing the edge $\{u,u'\}$.
            \item Case 4: $e = \{v'_*, v_*\}$. All shortest paths from $v'_*$ to any other vertex in $G$ must traverse $e$, and hence $e$ is monitored, e.g., by all pairs  $\{v'_*, u\}$ with $u \in VC$.
        \end{itemize}
\noindent The cardinality $|M|$ is equal to the sum of the cardinality of $M \setminus V(H)$ and the cardinality of $M \cap V(H)$ with $(M \setminus V(H)) \cap (M \cap V(H)) = \emptyset$. Since $|M \cap V(H)|$ is equal to the size $k$ of the \textsc{Vertex Cover} of the graph $H$ and $M \setminus V(H) = L'' \cup \{v_*'\}$, we can conclude that: \[ |M| = |M \cap V(H)| + |M \setminus V(H)| = k + |L''| + |\{v_*'\}| = k + |V(H)| + 1 \]
    \end{proof}
\end{lemma}

\newpage
\begin{lemma}
    \label{megToVertexCover}
    Let $M$ be a \textsc{MEG-set} of $G$ with size $k + |V(H)| + 1$ (for a suitable value of $k$). $M \cap V(H)$ is a \textsc{Vertex Cover} of $H$ with size at most $k$.
    \begin{proof}
        Let $M'= M \setminus (L' \cup \{v_*\})$ be the set obtained by removing the vertex $v_*$ and all the vertices of $L'$ from the \textsc{MEG-set} $M$. Due to Lemma \ref{superiorLeafLemma}, $M'$ is a \textsc{MEG-set} of $G$. Moreover since $(L' \cup \{v_*\}) \cap V(H) = \emptyset$, we can observe that $M' \cap V(H) = M \cap V(H)$.\\
        We prove that if $\{u,v\}$ is an edge of $H$ then at least one of its endpoints is in $M'$. \\ 
        By contradiction suppose that there exists some $\{u,v\} \in E(H)$ such that neither $u$ nor $v$ are in $M'$ and let $\{x,y\}$ be a pair of vertices in $M'$ that covers $\{u,v\}$. \\
        Notice that $\forall u,v \in V(G)$ there exists a path of length at most four that uses only edges in $E(G) \setminus E(H)$.  \\ This implies that the distance $d(x,y)$ from $x$ to $y$ in $G$ is at most $3$.\\
        Moreover $d(x,y) \ge 3 $ since otherwise we would have $\{x,y\} \cap \{u,v\} \neq \emptyset$ and $\{x,y\} \cap \{u,v\} \subseteq M'$. The above discussion implies that $d(x,y) = 3$ and, w.l.o.g., $x$ must be a neighbor of $u$ and $y$ must be a neighbor of $v$. \\ Let $N(u)$ and $N(v)$ be the set of neighbors of $u$ and $v$ in $G$, respectively.\\ We have that $N(u) \cap M' \subseteq V(H)$ (resp $N(V) \cap M' \subseteq V(H)$) since the only neighbor of $u$ (resp $v$) that is not in $V(H)$ is $u'$ (resp $v'$) but neither $u'$ nor $v'$ belong to $M'$. \\
        Therefore $x,y \in V(H)$ and using the fact that the diameter of $H$ is at most $2$, we have $d(x,y) \le \diam(H) \le 2$, which contradicts $d(x,y) \ge 3$. \\
        The above discussion shows that $M' \cap V(H)$ is a vertex cover of $H$. The size of such a vertex cover is $|M' \cap V(H)| \le  |M \cap V(H)|  =  |M| - |M \setminus V(H)| \le |M| - (|V(H)| + 1) = k$, where we used $(L'' \cup \{v'_*\}) \subseteq M \setminus V(H) $, as ensured by Lemma \ref{leafLemma}.
    \end{proof}
\end{lemma}
\noindent Now we are ready to give the main result of this section.
\begin{theorem}
\label{MegNPComplete}
    The \textsc{MEG-set} decision problem is $\mathcal{NP}$-Complete.
    \begin{proof}
For each graph $H$ with $\diam(H) \le 2$ it is possible to construct (in polynomial time) a graph $G$ as seen in this section. \\
From Lemmas \ref{vertexCoverToMeg} and \ref{megToVertexCover} we obtain that \\
      \mbox{$\exists$ \textsc{Vertex Cover} $VC$ of $H : |VC| \le k \iff \exists$ \textsc{MEG-set} $M$ of $G : |M| \le k + |V(H)| + 1$.} Since the \textsc{Vertex Cover} problem is $\mathcal{NP}$-Hard even for graphs $H$ with $\diam(H) \le 2$, this implies that the \textsc{MEG-set} decision problem is  $\mathcal{NP}$-Hard.\\
      Furthermore, we provide an algorithm, \textsc{MEG-check}, that runs in polynomial time and verifies whether $M \subseteq V(G)$ is a \textsc{MEG-set}. As $M$ is a certificate, the \textsc{MEG-set} decision problem is in $NP$. The algorithm \textsc{MEG-check} takes a graph $G$ and a subset $M \subseteq V(G)$ as input and returns True if and only if $M$ is a \textsc{MEG-set} of the graph $G$ and $|M| \le k$.
      \newpage
\noindent The \textsc{MEG-check} algorithm, given a graph $G$ and a subset of $M \subseteq V(G)$ of vertices, examines each pair of vertices $\{x,y\}$ in $[M]^2$ to determine the edges monitored by this pair. To check if an edge $e \in E(G)$ is monitored by a pair $\{u,v\}$ it suffices to remove the edge $e$ and check if the distance between $u$ and $v$ increases.
\begin{algorithm}
    \caption{MEG-check($G,M \subseteq V(G)$,$k$)}
    $S \gets \emptyset$\;
    \ForEach{$\{x,y\} \in [M]^2$}{
        $P_{xy} \gets $ Find a shortest path from $x$ to $y$ in $G$\;
        \For{each $e \in P_{xy}$}{
            $G' \gets (V(G),E(G) \setminus \{e\})$\;
            \If {$d_{G'}(x,y) > d_{G}(x,y)$}{
                $S \gets S \cup \{e\}$\;
            }
        }
    }
    \If {$S = E(G)$ \textbf{and} $|M| \le k$}{
        \Return True\;
    }
    \Return False\;
\end{algorithm}

\noindent The correctness of the algorithm stems from the observation that the distance from $x$ to $y$ increases when an edge $e \in E(G)$, monitored by vertices $x$ and $y$ in the subset $M$, is removed.
Suppose, for the sake of contradiction, that the removal of edge $e$ does not result in an increase in the distance from $x$ to $y$. This implies the existence of a shortest path between $x$ and $y$ that does not include $e$. Consequently, $e$ is not monitored by vertices $x$ and $y$ in $M$. However, this contradicts the assumption that $e$ is monitored by $x$ and $y$ in $M$. Thus, the algorithm ensures that the removal of monitored edges accurately reflects the changes in the shortest paths between vertices in the subset $M$.\\

\noindent The time complexity of the algorithm can be characterized as:
\begin{itemize}
	\item The outer loop iterates over each pair of vertices in $[M]^2$, resulting in $|[M]^2| = O(|V(G)|^2)$ iterations.
	\item Finding the shortest path $P_{xy}$ may involve running a shortest path algorithm that has a time complexity of $O(|E(G)|)$ since the graph is unweighted and connected.
	\item The inner loop iterates over the edges in the shortest path from $x$ to $y$ and performs operations inside. Notice that the number of edges contained in this shortest path must be in $O(|V(G)|)$, because a simple path from $x$ to $y$ contains at most $O(|V(G)|)$ edges.
    \item The check performed on the distances requires recalculating a shortest path which has a time complexity of $O(|E(G)|)$.
	\item The last \texttt{if} statement takes $O(|E(G)|)$ time.
\end{itemize}
\noindent The time complexity is $O(|V(G)|^2 \cdot (|E(G)| + (|V(G)| \cdot |E(G)|))) = O(|V(G)|^3 \cdot |E(G)|)$. Then the time complexity is polynomial in the size of the input.
\end{proof}
\end{theorem}

\newpage 
\section{Inapproximability results}
In the following, we use this notation:
\begin{itemize}
    \item $OPT_{vc}(G)$ denotes the size of the minimum \textsc{Vertex Cover} of the input graph $G$.
    \item $OPT_{meg}(G)$ denotes denote the size of the minimum \textsc{MEG-set} of the input graph $G$.
    \item $H_s$ denotes a simple, undirected graph that contains a special vertex $s$ and satisfies the following two properties: 
        \begin{enumerate}
            \item [i)] $H_s \setminus \{s\}$ has at least $3$ vertices and is connected;
            \item [ii)] $s$ is a universal vertex, i.e., $E(H_s)$ contains all edges $\{s,v\}$ with $v \in (V(H_s) \setminus \{s\})$.
        \end{enumerate}
    \item $H$ denotes a simple, undirected and connected graph such that $|V(H)| \ge 3$.
    \item $F$ denotes a simple, undirected and connected graph such that $|V(F)| \ge 3$ and $\diam{(F)} \le 2$.
\end{itemize}

\noindent In this section, we show that if there exists a polynomial-time $r$-approximation algorithm for the \textsc{MEG-set} \textit{Optimization Problem}, then it is possible to build a polynomial-time $(r+\epsilon)$-approximation algorithm for the \textsc{Vertex Cover} \textit{Optimization Problem}, where $\epsilon > 0$ is constant. To achieve this result we need to use a graph $G$ constructed by replicating a graph $H_s$. The idea is to modify the previous reduction (Section \ref{construction}) by creating $c \in \mathbb{N}^+$ copies of the graph $H_s$, in order to increment the number of vertices contained in the \textsc{MEG-set} of $G$ while preserving the size of the minimum \textsc{Vertex Cover} of $H_s \setminus \{s\}$. Notice that we construct a graph $H_s$ from a generic graph $H$ by adding a universal vertex $s$ to it. The size of the optimal \textsc{MEG-set} of $H_s$, i.e., $OPT_{vc}(H_s)$ is related to the size of the optimal \textsc{MEG-set} of $H$, i.e., $ OPT_{vc}(H)$. This will be useful to build an algorithm that $(r + \epsilon)$-approximates the \textsc{Vertex Cover} solution for an arbitrary graph $H$. Next, we present a construction that establishes a connection between the size of \textsc{Vertex Cover}s of $F$ and the size of \textsc{MEG-set}s of $G$, where $G$ is a graph created by replicating the graph $F$. This construction will be employed to replicate the graph $H_s$, which is constructed from $H$ and satisfies $\diam{(H_s)} \le 2, |V(H_s)| \ge 3$. \newpage

\noindent We denote the copy of the graph $F$ as $F_i$, with $V(F_i) = \{v_i : v \in V(F)\}$ and $E(F_i) =\{\{u_i,v_i\}\ : \{u,v\} \in E(F) \}$.\\
Initially, the graph $G = (\bigcup_{i=1}^c V(F_i),\bigcup_{i=1}^c E(F_i))$ contains $c$ copies of the graph $F$, where $c \in \mathbb{N}^+$ represents the number of copies of the original graph $F$.
Then for each vertex $v \in V(F)$ we add $2$ vertices $v'$ and $v''$ to $G$. Subsequently we connect $v_i$ to $v'$ and $v'$ to $v'' \quad \forall i \in \mathbb{N}: 1 \le i \le c$. Note that each vertex $v_i \in V(F_i)$ is connected to a single vertex $v'$. We denote the set containing all the vertices $v'$ as $L'$ and the set containing all the vertices $v''$ as $L''$. Note that $|L'| = |L''| = |V(F)|$. \\
\noindent Observe that: 
\begin{enumerate}
        \item[i)] each $v'' \in L''$ belongs to all \textsc{MEG-set}s of $G$ due to Lemma \ref{leafLemma};
        \item[ii)] each $v' \in L'$ does not belong to any optimal \textsc{MEG-set} of $G$ due to Lemma \ref{superiorLeafLemma}.
\end{enumerate}
\label{replication construction}
We denote the set of edges $\{v_i,v'\} \quad \forall i \in \mathbb{N}: 1 \le i \le c$ as $E_{L'}$ and the set of edges $\{v',v'' \}$ as $E_{L''}$. We shortcut all the shortest paths from one leaf to another: we insert two vertices $v_{*}$ and $v_{*}'$ and connect them together via the edge $\{v_*,v_*'\}$ , then we connect $v_{*}$ to all vertices $v' \in L'$. We denote the set containing all the edges $\{v_*,v'\}$, $v' \in L'$, as $E_*$. The vertex $v_{*}$ is used to create shortcuts between all shortest paths from a vertex in $L' \cup L''$ to a vertex in $L' \cup L''$. Notice that since $v_{*}'$ is a leaf it belongs to all \textsc{MEG-set}s of $G$ due to the Lemma \ref{leafLemma}. The vertex $v_{*}'$ is used to monitor all the edges in $E_* \cup \{v_{*},v_{*}'\}$.  \\~\\
\noindent More formally let $G = ((\bigcup_{i=1}^c V(F_i)) \cup L' \cup L'' \cup \{v_*,v_{*}'\},(\bigcup_{i=1}^c E(F_i)) \cup E_{L'} \cup E_{L''} \cup E_* \cup \{ \;  \{v_*,v_{*}'\} \; \}) $, where 
    \begin{itemize}
        \item $L' = \{v': v \in V(F)\}$
        \item $L'' = \{v'': v \in V(F)\}$
        \item $E_{L'} = \{\{v_i,v'\}: v \in V(F),i \in \mathbb{N}^+ \quad 1 \le i \le c\}$
        \item $E_{L''} = \{\{v'',v'\}:  v \in V(F)\}$
        \item $E_{*} = \{\{v_*,v'\}:  v \in V(F)\}$
    \end{itemize}
\noindent The following Figure \ref{replicationPicture} provides an example of construction of the graph $G$ with $c = 2$.
    \begin{figure}[H]
        \centering
        
        \begin{tikzpicture}            
                \node[shape=circle,draw=black,minimum size = 0.7cm] (a_1) at (0,0) {$a_1$};
                \node[shape=circle,draw=black,minimum size = 0.7cm] (b_1) at (2,0) {$b_1$};
                \node[shape=circle,draw=black,minimum size = 0.7cm] (c_1) at (2,2) {$c_1$};
                \node[shape=circle,draw=black,minimum size = 0.7cm] (d_1) at (0,2) {$d_1$};
                \node[shape=circle,draw=black,minimum size = 0.7cm] (a_2) at (4,0) {$a_2$};
                \node[shape=circle,draw=black,minimum size = 0.7cm] (b_2) at (6,0) {$b_2$};
                \node[shape=circle,draw=black,minimum size = 0.7cm] (c_2) at (6,2) {$c_2$};
                \node[shape=circle,draw=black,minimum size = 0.7cm] (d_2) at (4,2) {$d_2$};
                \node[shape=circle,draw=black,minimum size = 0.7cm,blue] (a') at (0,-2) {$a'$};
                \node[shape=circle,draw=black,minimum size = 0.7cm,blue] (b') at (2,-2) {$b'$};
                \node[shape=circle,draw=black,minimum size = 0.7cm,blue] (c') at (4,-2) {$c'$};
                \node[shape=circle,draw=black,minimum size = 0.7cm,blue] (d') at (6,-2) {$d'$};
                \node[shape=circle,draw=black,minimum size = 0.7cm,red] (a'') at (0,-4) {$a''$};
                \node[shape=circle,draw=black,minimum size = 0.7cm,red] (b'') at (2,-4) {$b''$};
                \node[shape=circle,draw=black,minimum size = 0.7cm,red] (c'') at (4,-4) {$c''$};
                \node[shape=circle,draw=black,minimum size = 0.7cm,red] (d'') at (6,-4) {$d''$};
                \node[shape=circle,draw=black,minimum size = 0.7cm,olive] (*) at (1,-6) {$v_*$};
                \node[shape=circle,draw=black,minimum size = 0.7cm,olive] (*') at (1,-7) {$v_*'$};
                \path [-,thick] (a_1) edge (b_1);
                \path [-,thick] (b_1) edge (c_1);
                \path [-,thick] (c_1) edge (d_1);
                \path [-,thick] (d_1) edge (a_1);
                \path [-,thick] (a_2) edge (b_2);
                \path [-,thick] (b_2) edge (c_2);
                \path [-,thick] (c_2) edge (d_2);
                \path [-,thick] (d_2) edge (a_2);
                \path [-,thick,blue] (a_1) edge (a');
                \path [-,thick,blue] (b_1) edge (b');
                \path [-,thick,blue] (c_1) edge (c');
                \path [-,thick,blue] (d_1) edge (d');
                \path [-,thick,blue] (a_2) edge (a');
                \path [-,thick,blue] (b_2) edge (b');
                \path [-,thick,blue] (c_2) edge (c');
                \path [-,thick,blue] (d_2) edge (d');
                \path [-,thick,red] (a') edge (a'');
                \path [-,thick,red] (b') edge (b'');
                \path [-,thick,red] (c') edge (c'');
                \path [-,thick,red] (d') edge (d'');
                \path [-,thick,olive,>=stealth, bend left=90] (c') edge (*);
                \path [-,thick,olive,>=stealth, bend right=-90] (d') edge (*);
                \path [-,thick,olive] (a') edge (*);
                \path [-,thick,olive] (b') edge (*);
                \path [-,thick,olive] (*') edge (*);
        \end{tikzpicture}
        \caption{An example of construction of the graph $G$ from the input graph $F = (\{a,b,c,d\},\{ \; \{a,b\}, \{b,c\}, \{c,d\}, \{d,a\} \; \})$. Vertices and edges replicated from the original graph are colored in black; vertices in $L'$ and edges in $E_{L'}$ are colored in \textcolor{blue}{blue}; vertices in $L''$ and edges in $E_{L''}$ are colored in \textcolor{red}{red}; vertices in $\{v_{*}',v_{*}\}$ and edges in $E_{*}$ are colored in \textcolor{olive}{olive}.} 
        \label{replicationPicture}
    \end{figure}
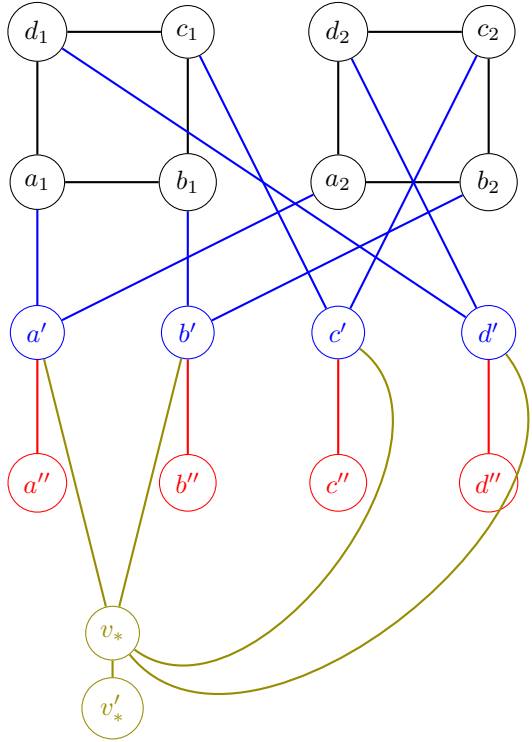
\newpage
\noindent In the following two lemmas, we prove that the existence of a \textsc{Vertex Cover} of $F$ with size $k$ implies the existence of a \textsc{MEG-set} of $G$ with size $ck + |V(F)| + 1$. Additionally, we prove the converse: if a \textsc{MEG-set} of $G$ with a size of $ck + |V(F)| + 1$ exists, then there exists a \textsc{Vertex Cover} of $F$ with size $k$.
To achieve this we use the same arguments used in Lemmas  \ref{vertexCoverToMeg} and \ref{megToVertexCover}.\\
We denote the number of copies of the graph $F$ in $G$ as $c \in \mathbb{N}^+$.
\begin{lemma}
\label{VertexCovertToMegReplicated}
Let $VC$ be a \textsc{Vertex Cover} of $F$ such that $|VC| \le k,$ $k\in \mathbb{N}^+$. There exists a \textsc{MEG-set} $M$ of $G$ such that $ |M| \le ck+|V(F)|+1$. 
\begin{proof}
To prove that this \textsc{MEG-set} exists we show $M$ explicitly. Let $VC_i$ be the set of vertices of $F_i$ corresponding to the vertices in $VC$: more formally $VC_i = \{v_i \in V(F_i): v \in VC\}$. This implies that $|VC_i| = |VC|$.
Consider the \textsc{MEG-set} $M = (\bigcup_{i=1}^{c} V(F_i)) \cup L'' \cup \{v_*'\}$. We can say that: \[|M| \le \left|\left(\bigcup_{i=1}^c VC_i\right) \cup L'' \cup \{v_*'\}\right| \le c|VC| + |L''| + |\{v_*'\}| = ck + |V(F)| + 1.\]
With the same arguments used in Lemma \ref{vertexCoverToMeg}, we can conclude that $M$ is a \textsc{MEG-set} of $G$, i.e., each edge of $G$ is monitored by at least a pair of vertices in $[M]^2$.
\end{proof} 
\end{lemma}

\begin{lemma}
\label{MegToVertexCoverReplicated}
Let $M$ be a \textsc{MEG-set} of $G$ such that $|M| \le ck + |V(F)| + 1$. There exists a \textsc{Vertex Cover} of $F$, say $VC$, of size $|VC| \le k$. Moreover, $VC$ can be found in polynomial time with respect to $|V(G)|$.
\begin{proof}
Let $M$ be a \textsc{MEG-set} of $G$ such that $|M| \le ck + n + 1$. Using Lemma \ref{leafLemma} we can say that $L'' \cup \{v_*'\} \subseteq M$. We denote the set obtained by removing all the leaves from the \textsc{MEG-set}  $M$ as $M' = M \setminus (L'' \cup \{v_*'\})$. Observe that \[|M'| \le |M| - |L'' \cup \{v_*'\}| \le (ck+|V(F)|+1) - (|V(F)|+1) = ck.\] We denote the set of all vertices contained in the set $V(F_i) \cap M'$ as $M_i'$. Notice that $M_\alpha'$ and $M_\beta'$, are disjoint, i.e.  $M_\alpha' \cap M_\beta' = \emptyset, $ $ 1 \le \alpha < \beta \le c$. Since $\sum_{i=1}^c |M_i'| \le |M'|$ and  $|M'| \le ck$ we can say that $\sum_{i=1}^c |M_i'| \le ck$. From the last inequality, we can observe that there exists $ h \in \mathbb{N}^+, 1 \le h \le c: |M_h'| \le k$. Using arguments similar to those in the proof of Lemma \ref{megToVertexCover} we can conclude that $M_h'$ is a \textsc{Vertex Cover} of $F_h$. This implies that $F$ contains a \textsc{Vertex Cover} of size $|F_h| \le k$. To find a \textsc{Vertex Cover} of size at most $k$ we must check at most $\Theta(c)$ copies of the graph $F$.
\end{proof}
\end{lemma}

\begin{theorem}
Let $A$ be a polynomial-time $r$-approximation algorithm for the \textsc{MEG-set} \textit{Optimization Problem} such that $r > 1$ is a constant value. \\ 
There exists a polynomial-time $(r+\epsilon)$-approximation algorithm for the \textsc{Vertex Cover} \textit{Optimization Problem}, where $\epsilon > 0$ is an arbitrarily small constant. 
\begin{proof}
We define an approximation algorithm called \textsc{Apx-VC} that takes a graph $H$ in input and returns an approximate solution for the \textsc{Vertex Cover} \textit{Optimization Problem} instance in input.
\begin{algorithm}
    \caption{\textsc{Apx-VC}($H$)}
    $VC_{b} \gets $ find, if it exists, a minimum size \textsc{Vertex Cover} of $H$ with less than $\frac{2 \cdot  r}{\epsilon}$ vertices\;
    $H_s \gets (V(H) \cup \{s\}, E(H) \cup \{ \{s,v\}: v \in V(H)\})$\;
    $G \gets$ graph constructed by replicating $H_s$ using the construction in Section \ref{replication construction} with $c = \left\lceil  \frac{2}{\epsilon}  \cdot (|V(H)|(r-1) + 2r - 2)  \right\rceil$\;
    $M_{apx} \gets A(G)$\;
    $VC_{s} \gets $ find a \textsc{Vertex Cover} of $H_s$ of size at most $\frac{|M_{apx}| - (|V(H_s)| + 1)}{c}$ as shown in Lemma~\ref{MegToVertexCoverReplicated}\;
    \If{$VC_{b}$ exists \textbf{and} $|VC_{b}| \le |VC_s|$}{
        $VC_{apx} = VC_{b}$ \;
    }
    \Else{
        $VC_{apx} = VC_s$\;
    }
    \Return $VC_{apx}$\;
\end{algorithm} \\
The algorithm \textsc{Apx-VC} is \textbf{correct} because both sets that can be returned, namely $VC_{s}$ and $ VC_{b}$, are \textsc{Vertex Cover}s of the graph $H$. By construction $VC_{b}$ is a \textsc{Vertex Cover} of $H$ and, due to Lemma \ref{MegToVertexCoverReplicated}, also $VC_{s}$ is a \textsc{Vertex Cover} of $H$. Moreover, the construction of the graph $G$ is well-defined since $\diam{(H_s) \le 2}$.

\noindent The algorithm \textsc{Apx-VC} takes \textbf{polynomial time} in the size of $V(H)$. To show this we analyze all the instructions for which it is not obvious to claim that they take polynomial time:
\begin{itemize}
    \item To find $VC_{b}$ we must check each \textsc{Vertex Cover} with size less than $\frac{2 \cdot  r}{\epsilon}$. The number of these \textsc{Vertex Cover}s is upper-bounded by the sum $\sum_{i=1}^{\left\lceil \frac{2 \cdot  r}{\epsilon} \right\rceil} |V(H)|^i$. This geometric sum is less than $|V(H)|^{\left\lceil \frac{2 \cdot  r}{\epsilon} \right\rceil + 1}$. Since $r$ is a constant value and $\epsilon > 0$ is a fixed constant we can conclude that finding $VC_{b}$ takes polynomial time in the size of $V(H)$.
    \item Since $r$ is a constant value and $\epsilon > 0$ is a fixed constant we observe that $c$ is a polynomial in the size of $V(H)$. This implies that the construction of the graph $G$ takes polynomial time in the size of $V(H)$.
    \item Due to Lemma \ref{MegToVertexCoverReplicated}, we can say that it is possible to find $VC_{s}$ in polynomial time in the size of $V(H)$.
\end{itemize}
First we can observe that $|V(H_s)| = |V(H)| + 1$, since we add the vertex $s$ to $V(H)$.
Due to Lemma \ref{VertexCovertToMegReplicated}, we can say that 
\begin{align*} 
OPT_{meg}(G) & \le c \cdot OPT_{vc}(H_s) + |V(H_s)| + 1 \\
& \le c \cdot OPT_{vc}(H_s) + (|V(H)| + 1) + 1 = c \cdot OPT_{vc}(H_s) + |V(H)| + 2.
\end{align*}

\noindent Combining the definition of  $r$-approximation algorithm and the last inequality we obtain: \[|M_{apx}| \le r \cdot OPT_{meg}(G) \le  r (c \cdot OPT_{vc}(H_s) + |V(H)| + 2).\]
By Lemma \ref{MegToVertexCoverReplicated}, we can conclude that:
\begin{align*}
|VC_{s}| & \le \frac{|M_{apx}| - |V(H_s)| - 1}{c} \le \frac{|M_{apx}|-|V(H)|-2}{c} \\
& \le \frac{r (c \cdot OPT_{vc}(H_s) + |V(H)| + 2) - (|V(H)| + 2)}{c}.
\end{align*}
We use the fact that $OPT_{vc}(H_s) \le OPT_{vc}(H) + 1$ to get the following:
\begin{align*}
    |VC_{s}| & \le r \cdot OPT_{vc}(H_s) + \frac{|V(H)|(r-1) + 2r - 2}{c} \\
    & \le r \cdot (OPT_{vc}(H)+1) + \frac{|V(H)|(r-1) + 2r - 2}{c}.
\end{align*}

\noindent By construction, if $OPT_{vc}(H) < \frac{2 \cdot r}{\epsilon}$, then $VC_{apx} = VC_{b}$ and $VC_{b}$ is the optimum solution. Then we suppose that $OPT_{vc}(H) \ge \frac{2 \cdot r}{\epsilon}$ . The solution returned by the algorithm has size at most $|VC_s \setminus \{s\}| \le |VC_s|$

We evaluate the approximation ratio of the found solution $VC_{s}$ using the previous inequalities:
\begin{align*}
    \frac{|VC_{s}|}{OPT_{vc}(H)} & \le r \cdot \frac{(OPT_{vc}(H)+1)}{OPT_{vc}(H)} + \frac{|V(H)|(r-1) + 2r - 2}{c \cdot OPT_{vc}(H)} \\
    & \le r + \frac{r}{OPT_{vc}(H)} + \frac{|V(H)|(r-1) + 2r - 2}{c \cdot OPT_{vc}(H)}.
\end{align*}
\noindent Since $OPT_{vc}(H) \ge \frac{2 \cdot r}{\epsilon}$ we deduce that $\frac{r}{OPT_{vc}(H)} \le \frac{\epsilon}{2}$. Furthermore, the number of copies is equal to $c =  \left\lceil  \frac{2}{\epsilon}  \cdot (|V(H)|(r-1) + 2r - 2)  \right\rceil$. This implies that 
\[\frac{|V(H)|(r-1) + 2r - 2}{c \cdot OPT_{vc}(H)} \le \frac{|V(H)|(r-1) + 2r - 2}{c} \le  \frac{\epsilon}{2}.\] Thus, combining the last $2$ inequalities we conclude that:
\[\frac{|VC_{s}|}{OPT_{vc}(H)} \le r + \frac{r}{OPT_{vc}(H)} + \frac{|V(H)|(r-1) + 2r - 2}{c \cdot OPT_{vc}(H)} \le r + \frac{\epsilon}{2} + \frac{\epsilon}{2} = r+\epsilon.\]
This means that if an algorithm approximates the \textsc{MEG-set} \textit{Optimization Problem} with a ratio of  $r$, the same algorithm can be used to approximate the \textsc{Vertex Cover} \textit{Optimization Problem} with an approximation ratio of  $(r+\epsilon)$.
\end{proof}
\end{theorem}

\newpage
\noindent Since the \textsc{Vertex Cover} \textit{Optimization Problem} is $\mathcal{APX}$-Hard \cite{10.1145/502090.502098}, we can conclude that the \textsc{MEG-set} \textit{Optimization Problem} is $\mathcal{APX}$-Hard. Otherwise, if the \textsc{MEG-set}
\textit{Optimization Problem} was not $\mathcal{APX}$-Hard it would have been possible to build an $(1 + \epsilon)$-approximation polynomial-time algorithm for the \textsc{Vertex Cover} \textit{Optimization Problem}.
\begin{corollary}
    The \textsc{MEG-set} \textit{Optimization Problem} is $\mathcal{APX}$-Hard.
\end{corollary}
\noindent If Unique Games Conjectures holds \cite{10.1145/509907.510017}, there is no polynomial-time $(2-\delta)$-approximation algorithm for the \textsc{Vertex Cover} \textit{Optimization Problem}, for any constant $\delta > 0$. We conclude that there does not exist a polynomial-time $(2-\delta)$-approximation algorithm for the \textsc{MEG-set} \textit{Optimization Problem} since a polynomial-time $(2-\delta)$-approximation algorithm for the \textsc{MEG-set} \textit{Optimization Problem}, would imply the existence of a polynomial-time $(2 - \delta + \epsilon)$-approximation algorithm for the \textsc{Vertex Cover} \textit{Optimization Problem}. Since we can choose $\epsilon \in  (0, \delta)$, this would contradict the unique games conjecture.
\begin{corollary}
    If the unique games conjecture holds then there exists no $(2-\delta)$-approximation algorithm for \textsc{MEG-set} \textit{Optimization Problem}, for any constant $\delta > 0$.
\end{corollary}

\newpage
\section{An efficient approximation algorithm}
In this section, we present a polynomial-time $O(\sqrt{|V(G)| \cdot \ln{|V(G)|}})$-approximation algorithm for the \textsc{MEG-set} \textit{Optimization Problem}. The algorithm is based on the well-known greedy $H_{n}$-approximation algorithm for the \textsc{Set Cover} \textit{Optimization Problem}, where $n$ is the size of the \textsc{Set Cover} instance universe and $H_{n} = \sum_{k=1}^n \frac{1}{k}$ is the $n$-th harmonic number. The algorithm idea is to transform a \textsc{MEG-set} instance, say $G$, into a \textsc{Set Cover} instance, whose universe $U$ is equal to the set of edges $E(G)$ and whose collection $S$ of subsets $S_{xy}$ of $U$ is such that, for each pair $\{x,y\}$ of vertices in $[V(G)]^2$, there is a subset that covers all edges $e \in U$ such that $e$ is monitored by the pair $\{x,y\}$. Since $U = E(G)$ by construction, the following algorithm calculates an $H_{|E(G)|}$-approximate solution with respect to the optimal number of pairs in $[V(G)]^2$ that are needed to monitor all edges. Then we show that this solution is $O(\sqrt{|V(G)| \cdot \ln{|V(G)|}})$-approximate with respect to the number of vertices in an optimal \textsc{MEG-set}. The following is the pseudo-code of the algorithm, namely \textsc{MEG-apx}, described above.
\begin{algorithm}
    \caption{\textsc{MEG-apx}($G$)}
    $S \gets \emptyset$\;
    \ForEach{$\{x,y\} \in [V(G)]^2$}{
        $S_{xy} \gets $ Find the edges in $E(G)$ monitored by pair $\{x,y\}$\;
        \If{$S_{xy} \neq \emptyset$}{
            $S \gets S \cup \{S_{xy}\}$\;
        }
    }
    $\mathcal{I} \gets $ A \textsc{Set Cover} instance where the universe is $U = E(G)$ and the subsets are in $S$\;
    $SC_{apx} \gets $ Find an $H_{|E(G)|}$-approximate solution for $\mathcal{I}$\;
    $M_{apx} \gets \emptyset$\;
    \ForEach{$S_{xy} \in SC_{apx}$}{
        $M_{apx} \gets M_{apx} \cup \{x,y\} $\; 
    }
    \Return $M_{apx}$\;
\end{algorithm} \\
\noindent $S$ is the set of subsets $S_{xy}$ containing the edges monitored by the pair $\{x,y\}$. \\
\noindent $U = E(G)$ is the set containing all the edges of the input graph $G$. \\
\noindent The \textbf{correctness} of the above algorithm derives from the following observation: if $SC_{apx}$ is a \textsc{Set Cover} of $\mathcal{I}$, then $M_{apx}$ is a \textsc{MEG-set} of $G$. Since $SC_{apx}$ is a \textsc{Set Cover}, all the edges of $E(G)$ are covered by the subsets in $SC_{apx}$. Due to the construction of the instance $\mathcal{I}$, each subset $S_{xy}$ contains all the edges monitored by the pair $\{x,y\}$. It follows that a valid \textsc{Set Cover} represents a subset of pairs in $[V(G)]^2$ such that every edge of $E(G)$ is monitored. Since all the vertices belonging to at least one of these pairs are in $M_{apx}$, all the edges in $E(G)$ are trivially monitored by the vertices in $M_{apx}$. \\~\\
To formally prove the correctness of the algorithm, suppose, for the sake of contradiction, that $M_{apx}$ is not a \textsc{MEG-set} of $G$. Consider the set $SC_{apx}$ from which $M_{apx}$ is constructed; $SC_{apx}$ must be a (not necessarily proper) subset of $[M_{apx}]^2$. Using the hypothesis that $M_{apx}$ is not a \textsc{MEG-set}, we have that there exists an edge $\{u,v\} \in E(G)$ not monitored by any pair in $[M_{apx}]^2$. Combining these propositions, we can conclude that there does not exist $S_{xy} \in SC_{apx}: \{x,y\} \in [M_{apx}]^2$ such that $S_{xy}$ covers $\{u,v\}$. Consequently, $SC_{apx}$ is not a \textsc{Set Cover} of $\mathcal{I}$, which contradicts the initial assumption. \\~\\
\noindent To show that the algorithm runs in \textbf{polynomial time} in the size of the input, we can observe that:
\begin{itemize}
    \item Checking if an edge $e$ is monitored by a pair of nodes $\{x,y\}$ takes $O(|E(G)|)$ time. To perform this operation, as seen in Section \ref{MegNPComplete}, it is sufficient to verify whether, after the removal of edge $e$, the distance between $x$ and $y$ increases. 
    \item Due to the previous statement, generating $\mathcal{I}$ takes $O(|V(G)|^2 \cdot |E(G)|)$ time.
    \item $SC_{apx}$ is found, using the greedy algorithm in polynomial time, with a simple implementation that takes $O(|V(G)|^6)$ time.
    \item  $M_{apx}$ is constructed iterating all subsets $S_{xy}$ of $S$, this takes $O(|V(G)|^2)$ time.
\end{itemize}
\noindent Combining all these statements, we deduce that the \textsc{MEG-apx} algorithm runs in polynomial time in the size of the input. \\
In the following, we show that the \textbf{approximation ratio} of the \textsc{MEG-apx} algorithm is $O(\sqrt{|V(G)| \cdot \ln{|V(G)|}})$. Let us recall that $H_{n} = O(\ln{(n)})$, we use this assumption without explicitly referencing it in the following proof.
\begin{theorem}
\textsc{MEG-apx} is an $O(\sqrt{|V(G)| \cdot \ln{|V(G)|}})$-approximation algorithm.     \begin{proof}
        We denote the \textsc{MEG-set} with minimum size as $M^*$ and the solution found by the algorithm \textsc{MEG-apx} as $M_{apx}$.
        Let $S$ be the collection of sets computed by \textsc{MEG-apx}, i.e., $S$ contains all sets $S_{xy}$ such that $\{x,y\}$ monitors at least one edge. Since, by definition, each edge of $G$ is monitored by at least a pair in $[M^*]^2$, $S$ is a \textsc{Set Cover} of the instance $\mathcal{I}$. Let $C^*$ be the size of the optimal solution for instance $\mathcal{I}$. We can observe that $C^* \le |S| \le |M^*|^2$. Since $M_{apx}$ is constructed from $SC_{apx}$, we can deduce that the size of $M_{apx}$ is at most $2 \cdot |SC_{apx}|$. By definition of $r$-approximation algorithm, the size of $SC_{apx}$ is at most $O(\ln{|E(G)|}) \cdot C^*$. The previous inequalities imply the following: 
        \begin{align*}    
         |M_{apx}| &\le 2 \cdot |SC_{apx}| \le 2 \cdot O(\ln{|E(G)|}) \cdot C^* = O(\ln{|E(G)|}) \cdot C^*\\
         &= O(\ln{|V(G)|^2}) \cdot C^* = O(\ln{|V(G)|}) \cdot C^* \le O(\ln{|V(G)|}) \cdot |M^*|^2\\
         &= O( |M^*|^2 \cdot \ln{|V(G)|}). 
         \end{align*}
Using the last inequality we can show the following upper bound on the approximation ratio of the algorithm.
\[ \frac{|M_{apx}|}{|M^*|} \le \frac{O(|M^*|^2 \cdot \ln{|V(G)|})}{|M^*|} = O(|M^*| \cdot \ln{|V(G)|}).\] 
Since $|M_{apx}|$ is trivially at most $|V(G)|$, we can obtain another upper bound on the approximation ratio.
\[ \frac{|M_{apx}|}{|M^*|} \le \frac{|V(G)|}{|M^*|}. \]
Finally, by combining the two upper bounds we can deduce that:
\[ \frac{|M_{apx}|}{|M^*|} = O\left( \min{\left\{|M^*| \cdot \ln{|V(G)|}, \frac{|V(G)|}{|M^*|}\right\}}\right).\]
We can conclude that the previous minimum is maximized (Figure \ref{minimizationGraph}) when $|M^*| = \sqrt{\frac{|V(G)|}{\ln{|V(G)|}}}$, This implies that:
\[ \frac{|M_{apx}|}{|M^*|} = O\left(\min{\left\{|M^*| \cdot \ln{|V(G)|}, \frac{|V(G)|}{|M^*|}\right\}}\right) = O\left(\sqrt{|V(G)| \cdot \ln{|V(G)|}}\right).\]
\end{proof}

\begin{figure}[H]
        \centering
        
    \begin{tikzpicture}
        \begin{axis}[
            xlabel={$m^*$},
            ylabel={},
            grid=both,
            legend pos=north west,
            xmin=0,
            xmax=15,
            ymin=0,
            ymax=15,
        ]
        
        \addplot[blue, domain=0.1:10, samples=100,very thick] {5/x};
        \addlegendentry{$\frac{5}{m^*}$}
        
        \addplot[red, domain=0.1:10, samples=100,very thick] {5 * ln(x)};
        \addlegendentry{$m^* \ln(5)$}
    
        \addplot[green, domain=0.1:10, samples=100,dashed, very thick] {min(5/x, 5 * ln(x))};
         \addlegendentry{$\min{\{m^* \ln(5), \frac{5}{m^*}}\}$}
         
        \end{axis}
    \end{tikzpicture}
    \caption{Example of $\min{\{|M^*| \cdot \ln{|V(G)|}, \frac{|V(G)|}{|M^*|}\}}$ as a function of $|M^*|$, when $|V(G)|=5$.}
    \label{minimizationGraph}
\end{figure}
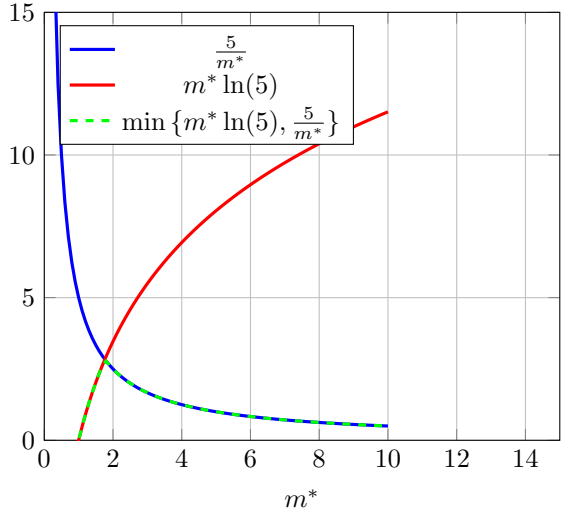

\end{theorem}

\newpage

\section{Conclusion}
In this thesis, we studied the \textsc{MEG-set} \textit{Optimization problem} that involves determining the minimum set of vertices that need to be monitored to achieve geodetic coverage for a given network to ensure complete coverage of edges and mitigate faults by monitoring lines of communication.
The problem has been recently introduced \cite{10.1007/978-3-031-25211-2_19} and has intriguing practical implications.
We have provided new results on both the positive and negative aspects. On the negative side, we have found a reduction from \textsc{Vertex Cover}, showing that the problem is $\mathcal{APX}$-Hard and that, if the unique games conjecture holds, the problem is non-approximable within a factor of $2-\epsilon$, for any constant $\epsilon>0$.
On the positive side, we have found a polynomial-time $\sqrt{|V(G)| \cdot \ln{|V(G)|}}$-approximation algorithm, where $|V(G)|$ represents the size of the set of vertices in the graph $G$. Possible future developments may focus on strengthening the inapproximability result, perhaps by providing a reduction from the \textsc{Set Cover} problem. To this aim, it might be useful to employ the graph replication strategy described in this thesis.

\newpage

\bibliographystyle{plain} 
\bibliography{references} 

\end{document}